\documentclass[fleqn,11pt,twoside]{article}

\usepackage{amsthm,amsthm,amssymb, color, xcolor,epsfig, graphics, subfigure}

\usepackage{amsmath, graphicx, latexsym, lscape }

\makeatletter
\newcommand{\copyrightnote}[2]{{\renewcommand{\thefootnote}{}
 \footnotetext{\small\it
\begin{flushleft}
 \copyright \ #1   #2  
\end{flushleft}}}}

\newcommand{\Name}[1]{\begin{flushleft}
                       \LARGE \bf #1
                       \end{flushleft}\vspace{-3mm}}

\newcommand{\Author}[1]{\begin{flushleft}
                       \it #1 \end{flushleft}}

\newcommand{\Address}[1]{\begin{flushleft}
                       \it #1 \end{flushleft}}

\newcommand{\Date}[1]{\begin{flushleft}
                      \small  \it #1 \end{flushleft}}

\newcommand{\evenhead}{Author \ name}
\newcommand{\oddhead}{Article \ name}

\renewcommand{\@evenhead}{
\hspace*{-3pt}\raisebox{-15pt}[\headheight][0pt]{\vbox{\hbox to \textwidth
{\thepage \hfil \evenhead}\vskip4pt \hrule}}}
\renewcommand{\@oddhead}{
\hspace*{-3pt}\raisebox{-15pt}[\headheight][0pt]{\vbox{\hbox to \textwidth
{\oddhead \hfil \thepage}\vskip4pt\hrule}}}
\renewcommand{\@evenfoot}{}
\renewcommand{\@oddfoot}{}

\long\def\@makecaption#1#2{%
  \vskip\abovecaptionskip
  \sbox\@tempboxa{\small \textbf{#1.}\ \ #2}%
  \ifdim \wd\@tempboxa >\hsize
    {\small \textbf{#1.}\ \ #2}\par
  \else
    \global \@minipagefalse
    \hb@xt@\hsize{\hfil\box\@tempboxa\hfil}%
  \fi
  \vskip\belowcaptionskip}

\newcommand{\JNMPnumberwithin}[3][\arabic]{%
  \@ifundefined{c@#2}{\@nocounterr{#2}}{%
    \@ifundefined{c@#3}{\@nocnterr{#3}}{%
      \@addtoreset{#2}{#3}%
      \@xp\xdef\csname the#2\endcsname{%
        \@xp\@nx\csname the#3\endcsname .\@nx#1{#2}}}}%
}

\newcommand{\resetfootnoterule} {
  \renewcommand\footnoterule{%
  \kern-3\p@
  \hrule\@width.4\columnwidth
  \kern2.6\p@}
}

\renewcommand{\footnoterule}{}

\makeatother

\theoremstyle{definition}

\begin{document}

\renewcommand{\evenhead}{
{\LARGE\textcolor{blue!10!black!40!green}{{\sf \ \ \
]ocnmp[}}}\strut\hfill H. Aratyn, J.F. Gomes, A.H. Zimerman}
\renewcommand{\oddhead}{ {\LARGE\textcolor{blue!10!black!40!green}{{\sf ]ocnmp[}}}\ \ \ \ \   
Dressing Chains of Even Periodicity}

%%%% Matter for the first page 
\thispagestyle{empty}
\newcommand{\FistPageHead}[3]{
\begin{flushleft}
\raisebox{8mm}[0pt][0pt]
{\footnotesize \sf
\parbox{150mm}{{Open Communications in Nonlinear Mathematical Physics}\ \  \ {\LARGE\textcolor{blue!10!black!40!green}{]ocnmp[}}
\ \ Vol.2 (2022) pp
#2\hfill {\sc #3}}}\vspace{-13mm}
\end{flushleft}}

\FistPageHead{1}{\pageref{firstpage}--\pageref{lastpage}}{ \ \ Article}

\strut\hfill

\strut\hfill

\copyrightnote{The author(s). Distributed under a Creative Commons Attribution 4.0 International License}

\Name{On Hamiltonian Formalism for  Dressing Chain
Equations of Even Periodicity}

\Author{H. Aratyn$^{\,1}$, J.F. Gomes$^{\,2}$ 
and A.H. Zimerman$^{\,2}$}

\Address{$^{1}$ Department of Physics, 
University of Illinois at Chicago, 
845 W. Taylor St.
Chicago, Illinois 60607-7059, USA\\[2mm]
$^{2}$ Instituto de F\'{\i}sica Te\'{o}rica-UNESP,
Rua Dr Bento Teobaldo Ferraz 271, Bloco II,
01140-070 S\~{a}o Paulo, Brazil }

\Date{Received: October 17, 2022; Accepted: November 8, 2022}

\setcounter{equation}{0}

\begin{abstract}
\noindent 
We propose a Hamiltonian formalism for $N$ periodic dressing chain 
with  the even number $N$. The formalism is based on Dirac reduction 
applied to the $N+1$ periodic  dressing chain with the odd number $N+1$ for 
which the Hamiltonian formalism is well known. 
The Hamilton dressing chain equations in the $N$ even case 
depend explicitly on a pair of conjugated Dirac constraints and are
equivalent to $A^{(1)}_{N-1}$ invariant
symmetric  Painlev\'e equations.
\end{abstract}

\label{firstpage}

%%%% The Article text starts here

\section{Introduction}
\label{section:intro}
The $N$ periodic dressing chain has emerged in a study 
of Schr\"odinger operators interconnected
by Darboux transformations \cite{veselov}. Its
equivalence to $A^{(1)}_{N-1}$ invariant
Painlev\'e equations  has been subject of  a number of
papers, see e.g. \cite{adler,SHC06,takasaki,Tsuda04,WH03}.
The system can  naturally be realized  as a self-similarity limit 
of the $t_2$ flow equations of several two-dimensional integrable
models, 
e.g. (1) the integrable mKdV hierarchy on the loop algebra of 
$ sl(N)=A_{N-1}$ endowed with a principal gradation 
\cite{schiff,victoretal}, (2) a class of  constrained KP
hierarchy referred to as $2n$-boson  integrable models generalizing 
AKNS hierarchy \cite{2010paper,2011paper} that in a self-similarity
limit are equivalent to the  $N=2n+1$ periodic dressing chain equations.

The Hamiltonian formalism for $N$  periodic dressing chain \cite{veselov}
is straightforward as long as $N$ is odd.
However the same structure 
for even $N$  requires for consistency  an additional relation 
\cite{takasaki}.
Here, we propose a coherent formalism based on Dirac approach where the
consistency condition naturally emerges as a secondary constraint for even $N$
and does not need to be set to zero.
We illustrate our formalism by explicitly 
employing the reduction from the Hamiltonian formalism  of $N=5$ 
to $N=4$ dressing chains. 
We show equivalence between the $N=4$ dressing chain equations
with  explicit dependence on two Dirac constraints
and  the $A^{(1)}_3$ symmetric Painlev\'e V equations.

Furthermore, we observe in this paper that generalization of the method
to  $2n+1 \to 2n$ reductions for  $n$ integer and $n \ge 2$ follows the same steps
and we provide a formula describing all such Hamilton equations. 
We establish the invariance of the two Dirac  constraints %$\Phi,\Psi$  
under extended affine $A^{(1)}_{2n-1}$ 
Weyl symmetry  that ensures invariance of the dressing chain equations.
Moreover the $A^{(1)}_{2n-1}$ basic Weyl algebra relations hold 
despite the presence
of additional (Dirac) terms  in fundamental brackets between  currents $j_i,
i=1,{\ldots} , 2n$ and the Hamiltonian. 

The paper starts by revisiting basic facts about periodic dressing
chain equations and the corresponding Hamiltonian formalism in section
\ref{section:background}. We stress a different nature of commutation
relations satisfied by the quantity $ \sum_{i=1}^N j_i$ with
respect to the basic Poisson bracket $\{ \cdot,\cdot\}$ for $N$ odd
and even.

The main results of the paper are presented in section \ref{section:reduction}
and derived from Dirac reduction procedure in subsection \ref{section:hamdir}.
The subsection \ref{section:invariance} is devoted to proving that 
despite explicit presence of Dirac constraints in Hamilton equation of
$N$ periodic dressing chain equations they remain invariant under
extended affine Weyl symmetry group $A^{(1)}_{N-1}$.

We describe a mechanism to deform the  $A^{(1)}_{4}$ symmetric
Painlev\'e equations in section \ref{section:deformation}.
The reduction formalism $A^{(1)}_{4} \to A^{(1)}_{3}$ we have introduced
offers new ways to explore how the  $A^{(1)}_3$ Weyl symmetry can be 
broken down to a partial B\"acklund symmetry by addition of quadratic
terms to the Hamiltonian of $N=5$ symmetric Painlev\'e equations. 

%%%%%%%%%%

In the case of  $2n$-boson constrained KP
hierarchy \cite{2010paper,2011paper} the self-similarity reduction
yields Hamiltonians of the  $A^{(1)}_{2n}$ Painlev\'e models 
expressed in terms of $n$ canonical pairs 
$e_i, Y_i, i=1,2, {\ldots} , n$ satisfying brackets 
$\{ e_i , Y_j \} = \delta_{ij}, \; i=1, {\ldots}, n$.
The Dirac formalism was defined in 
\cite{2011paper} to further reduce this system to 
$A^{(1)}_{2n-1}$ Painlev\'e models by setting $Y_n=0$. 
This construction was illustrated by reduction of $A^{(1)}_{4}$ 
Painlev\'e equations down to the $A^{(1)}_{3}$ symmetric Painlev\'e V equations.
Although the reference \cite{2011paper} employed a different set of 
variables than the conventional periodic dressing chain
variables $j_i, i=1,{\ldots} ,2n+1$ of this paper 
in section \ref{section:2011paper} we are able to show 
that the reduction of \cite{2011paper} is in complete agreement with the 
framework put forward in this paper.
The advantage of the current formalism is that is more intuitive and
therefore simpler to generalize from the $n=2$ four boson model to 
higher $n$ models.

\section{Background}
\label{section:background}
Our starting point is the set of  $N$  periodic 
dressing chain equations:
\begin{equation}
\left( j_i +j_{i+1} \right)_z =  
- \left( j_i -j_{i+1} \right)
\left( j_i +j_{i+1}  \right) + \alpha_i
, \qquad
i=1,{\ldots} ,N\, ,
\label{mape}
\end{equation}
where periodicity implies that $j_{N+i}=j_i, \alpha_{N+i}=\alpha_{i},
i=1,{\ldots}, N-1$.  

Let us first work with the odd number $N=2n+1$ and recall 
the Hamiltonian formalism  for such case \cite{veselov}. 
Note that for the odd $N$ one can invert the relation
\begin{equation}
f_i=j_i+j_{i+1}\, ,
\label{ffromji}
\end{equation}
 by providing $j_i$ in terms of  $f_k, k=1,{\ldots} ,N$ :
\begin{equation}
j_i = \frac12 \sum_{k=0}^{N-1} (-1)^k f_{i+k} \, .
\label{fromf2j}
\end{equation}
For example for $N=3$:
\[ j_1 =\frac12 ( f_1-f_2+f_3), \;  j_2 = \frac12 ( f_2-f_3+f_1), \; 
j_3  = \frac12 ( f_3-f_1+f_2)\,.
\]

Thus as long as $N$ is odd one is able to invert relation \eqref{ffromji}
and pass effortlessly from $f_i$-formalism to
$j_i$-formalism establishing full equivalence 
between the bracket structure 
\begin{equation}
\{ f_i , f_{k} \}= \begin{cases} 0 & k \ne i+1, k+1 \ne i \\
1 & k=i+1\\
-1& k+1=i\, ,
\end{cases}
\label{fifk}
\end{equation}
and the bracket structure:
\begin{equation}
\{ j_i , j_k \}= (-1)^{k-i+1}, \, 1 \le i < k \le N \, . %\{ j_i , j_i \}=0, \; 
\label{Noddpb}
\end{equation}
%and vice-versa.

For a sum $\sum_{k=1}^N j_k$ it holds from \eqref{Noddpb} that 
\begin{equation}
\{ j_i, j_1+j_2+j_3+j_4+j_5+{\ldots} +j_N \} = \begin{cases}
(-1)^{i+1}& N\;\;{\rm even}\\
0& N\;\;{\rm odd}
\end{cases}\, .
\label{JiJN}
\end{equation}
Thus setting  the quantity $\Phi=\sum_{k=1}^N j_k$ to be equal to a constant 
is consistent with the underlying Poisson bracket structure for
$N$-odd but  for $N$-even it will be imposed below through the Dirac system of
primary/secondary constraints.

Defining,  like Shabat and Veselov in \cite{veselov}, the cubic Hamiltonian
\begin{equation}
H_{N} = \frac13 \sum_{k=1}^N j_k^3 + \sum_{k=1}^N \beta_k j_k\, ,
\label{Hjbeta}
\end{equation}
one obtains
\begin{equation}
\left\{ j_i+ j_{i+1} \, ,\, H_{N} \right\}= j_{i+1}^2-j_{i}^2
+\beta_{i+1}-\beta_{i}, \quad i=1,2,{\ldots} , N=2n+1\, ,
\label{jijip1HV}
\end{equation}
using the bracket \eqref{Noddpb} for $N$ being odd.
We recognize in equations \eqref{jijip1HV}
the dressing chain equations \eqref{mape} with
\begin{equation}
\alpha_i= \beta_{i+1}-\beta_{i},\qquad i=1,{\ldots} ,N \, .
\label{alphabetai}
\end{equation}
The above relation implies that $\sum_{k=1}^N \alpha_k=0$.

Identifying in the Hamilton equations 
the expression $\{ f, H\}$ with  $f_z$ makes it possible 
to allow for $\sum_{k=1}^N \alpha_k\ne 0$ 
by redefining $j_i$ by e.g. $ J_i= j_i - \frac14 z$ (see e.g. \cite{veselov,victoretal})
and working with the corresponding Hamiltonian  :
\begin{equation} 
H_{N} = \frac13 \sum_{k=1}^N J_k^3 + \frac{z}{4} \sum_{k=1}^N J_k^2
+ \sum_{k=1}^N \beta_k J_k \, .
\label{HamJ}
\end{equation}
For simplicity we will continue to work here with the Hamiltonian \eqref{Hjbeta}.

\section{Reduction from $N+1$ periodic dressing chain to even $N$  
periodic dressing chain}
\label{section:reduction}

The main result of the paper is development of the Hamiltonian formalism
for even $N=2n$ case of equation
\eqref{jijip1HV}, which in general case enters the Hamiltonian formalism 
in the following form :
\begin{equation} 
\{ j_i+j_{i+1} ,H \} = j_{i+1}^2-j_i^2+\beta_{i+1} -\beta_i
+(-1)^{i+1} \frac{(j_i+j_{i+1}) }{\Phi}\Psi, \quad i=1,2,{\ldots} ,
N=2n \, ,
\label{main}
\end{equation}
with $\Phi \ne 0, \Psi \ne 0$, where $\Phi$ and $\Psi$ are defined as
\begin{equation} 
\Phi = \sum_{k=1}^{N=2n} j_k, \quad \Psi= 
\sum_{k=1}^{N=2n} (-1)^{k+1} (j_k^2 +\beta_k) \, .
\label{PhiPsi}
\end{equation}
The Hamiltonian $H$ in relation \eqref{main} will be
derived below from the Dirac formulation and the bracket $\{
\cdot,\cdot\}$ always refers to the Poisson bracket \eqref{Noddpb}.

We can cast the equations \eqref{main} into 
symmetric $A^{(1)}_{N-1}$ Painlev\'e  equations. In particular
for $N=4$ we obtain $A^{(1)}_{3}$ symmetric Painlev\'e V equations:
\begin{equation}
\begin{split}
\Phi \, \{ f_1 ,H \} &=f_1 f_3 \left(f_2-f_4\right) -
\left(\alpha_1+\alpha_3\right) f_1 + \alpha_1 (f_1+f_3) \, ,\\
\Phi \,\{ f_2 ,H \}&=f_2 f_4 \left(f_3-f_1\right) +\left(\alpha_1+\alpha_3
\right) f_2 + \alpha_2 (f_2+f_4) \, ,\\
\Phi\, \{ f_3 ,H \}&=f_1 f_3 \left(f_4-f_2\right) -
\left(\alpha_1+\alpha_3\right) f_3 + \alpha_3 (f_1+f_3) \, ,\\
\Phi\, \{ f_4 ,H \}&=f_4 f_2 \left(f_1-f_3\right) +
\left(\alpha_1+\alpha_3 \right) f_4 + \alpha_4 (f_2+f_4) \, ,
\label{N4feqs}
\end{split}
\end{equation}
as follows by verification after substitution of $f_i, \alpha_i$ from 
equations \eqref{ffromji}, \eqref{alphabetai}. Remarkably after 
inserting values $\Phi, \Psi$ from the definition \eqref{PhiPsi} the right
hand sides of equations \eqref{main}  can be expressed entirely by variables $f_i,
i=1,{\ldots} ,N$ without any need to invert relations \eqref{ffromji}. 
This observation is crucial for consistency of the proposed Hamiltonian 
formalism for even $N$ and arbitrary non-zero $\Psi$.

\subsection{Defining Hamiltonian formalism in terms of Dirac reduction}
\label{section:hamdir}

In section \ref{section:background} 
we have been working with the Hamiltonian system \eqref{jijip1HV} 
for odd $N=2n+1$ with $\Phi \equiv \sum_{k=1}^{N=2n+1} j_k $ that commutes with all 
$j_i$ and $H_N$ and can naturally be chosen to be a constant.
We will show how to obtain the same result for 
$\Phi \equiv \sum_{k=1}^{N} j_k $ with $N$ even by working with Dirac
formalism.

We now present the Dirac reduction formalism leading  
to the dressing chain equation \eqref{main}.  For illustration we perform
all the steps for the $N=5 \to N=4$ case. 

The first step is to eliminate $j_5$ from equation  the initial cubic 
Hamiltonian \eqref{Hjbeta}
expressing it in terms of the remaining four
objects $j_i, i=1,2,3,4$.
Elimination of $j_5$ 
involves setting the condition:
\begin{equation}
\psi_1=j_5=-j_1-j_2-j_3-j_4 \sim {\rm const} \, .
\label{psi1}
\end{equation}
 In this limit the Hamiltonian $H_{N=5}$ becomes
\begin{equation}
 \begin{split}
 H_{R}&= H_{N} \vert_{j_5 = - j_1- j_2- j_3- j_4}= 
 \frac13 \sum_{i=1}^4 j_i^3 -\frac13 (j_1+j_2+j_3+j_4)^3\\
&+ \sum_{i=1}^4 \beta_i j_i -\beta_5 (j_1+j_2+j_3+j_4)\,.
 \label{HVRj}
 \end{split}
 \end{equation}
Condition \eqref{psi1} needs to be accompanied by secondary Dirac condition
\begin{equation}
\psi_2= \Psi=  \{ j_5, H_{N=5}\}= \{ \psi_1, H_{N=5}\}
= - \sum_{k=1}^4 \left( (-1)^k j_k^2 
%+ \frac{z}{2} (-1)^k J_k 
+ \beta_k (-1)^k
\right) \, .
\label{j5zzero}
 \end{equation}
$\Psi$ can also be rewritten as 
\begin{equation}
 \Psi=  \left( j_1^2+j_3^2-j_2^2-j_4^2 \right)  
+\frac12 (-\alpha_1 +\alpha_2-\alpha_3+\alpha_4 )\, ,
\label{j5zzero1}
\end{equation}
in terms of $\alpha$'s.
The Dirac constraints $\psi_1 = -\Phi,\psi_2=\Psi$ satisfy the bracket
\begin{equation}
\{ \psi_1, \psi_2 \} = 2 (j_1+j_2+j_3+j_4)= 2 \Phi\, ,
\label{psi1psi2}
\end{equation}
that for $\Phi \ne 0$ gives rise to a regular matrix
\begin{equation}
D_{\alpha \beta}= \{ \psi_\alpha, \psi_\beta \}= 
\left( \begin{matrix} 0& 2 \Phi \\
-2 \Phi & 0
\end{matrix} \right), \quad D^{-1}_{\alpha \beta}=\frac{1}{2 \Phi}
\left( \begin{matrix} 0& -1 \\
1 & 0
\end{matrix} \right)\, ,
\label{Dmatrix}
\end{equation}
that will be used to calculate the Dirac bracket:
\begin{equation}
\{ j_i , j_k \}_D =\{ j_i , j_k \}-\{ j_i , \psi_\alpha \}\, D^{-1}_{\alpha \beta}
\{  \psi_\beta, j_k \}\, .
\label{Diracb1}
\end{equation}

Let us calculate $\{ j_1 , j_2 \}_D$ as illustration of the Dirac
bracket technique in this context.
{}Using  brackets:
\[ %\begin{split}
\{ j_1 , \psi_1 \}= - \{ j_1 , \Phi \}= -1,\;
\{  \psi_2, j_2 \}= 2(j_1-j_3-j_4).\;
\{ j_1 , \psi_2 \}= - 2 (j_2+j_3+j_4),\;
\{  \psi_1, j_2 \}= -1\, ,
%\end{split}
\]
we find 
\begin{equation}\begin{split}
\{ j_1 , j_2 \}_D &=1 - \{ j_1 , \psi_1 \} D^{-1}_{12}
\{  \psi_2, j_2 \} - \{ j_1 , \psi_2 \} D^{-1}_{21}
\{  \psi_1, j_2 \} =1 -\frac{j_1-j_3-j_4}{\Phi}\\&
- \frac{j_2+j_3+j_4}{\Phi}=\frac{\Phi}{\Phi} 
-\frac{j_1+j_2}{\Phi}
=\frac{j_3+j_4}{\Phi}=\frac{f_3}{\Phi}\, .
\label{diracbj1j2}
\end{split}
\end{equation}
Proceeding in same manner with calculations for all the other
non-vanishing brackets 
$\{j_i, j_j \}_D$
one obtains
\begin{equation}
\begin{split}
\{ j_1, j_4 \}_D &= -\frac{f_2}{\Phi}, \quad  \{ j_3, j_4
\}_D=\frac{f_1}{\Phi},\quad \{ j_1, j_2 \}_D = \frac{f_3}{\Phi}, \\  
\{ j_2, j_4 \}_D&=-\frac{f_4}{\Phi}+\frac{f_3}{\Phi} , \quad
\{ j_2, j_3 \}_D = \frac{f_4}{\Phi}, \quad  \{ j_1, j_3 \}_D=-\frac{f_3}{\Phi}
+\frac{f_2}{\Phi}
\label{taksolutions}
\end{split}
\end{equation}
and it follows indeed that $\{ j_i, \Phi\}_D=0$ for any $i$.

Takasaki in \cite{takasaki} obtained the  brackets \eqref{taksolutions} 
by first assuming that the periodic dressing chain equations \eqref{mape} 
hold for $N=4$
and then inserting equations \eqref{mape} into the alternating sum 
$\sum_{k=1}^{N} (-1 )^k \left(j_k+j_{k+1}\right)_z$  that is
identically zero as long as  $N$ is even.
Imposing consistency of those two equations amounts to setting $\Psi=0$.
This in turn makes it possible to invert equations \eqref{ffromji}
and find expressions for  $ j_i$ in terms of $f_i$.
The brackets $\{j_i,j_j\}$ can in such way be derived 
from brackets \eqref{fifk} for $f_i$
leading to an alternative derivation of  \eqref{taksolutions} \cite{takasaki}.

Here we will instead follow  the Dirac procedure 
and use the original Poisson bracket $\{ j_i, j_k\}=
(-1)^{k-i+1}$ with the Hamiltonian:
\begin{equation}
H= H_{R}+ \lambda_1 \psi_1+\lambda_2 \psi_2\, ,
 \label{DiracHam}
\end{equation}
where $H_{R}$ is given in 
\eqref{HVRj} and $\lambda_\alpha, \alpha=1,2$ are Lagrange multipliers that can be
derived from
\begin{equation}
%\begin{split}
0= \{ \psi_\alpha, H_{R} \} + \sum_{\beta=1}^2 \lambda_\beta \{ \psi_\alpha, \psi_\beta
\}= \{ \psi_\alpha, H_{R} \} + \sum_{\beta=1}^2  
D_{\alpha \, \beta}\lambda_\beta \, ,
\label{Dconst1}
%\end{split}
\end{equation}
or 
\[
\lambda_\alpha= -  \sum_{\beta=1}^2 D^{-1}_{\alpha \, \beta} \{
\psi_\beta, H_{R} \}\;\to\; 
\lambda_1= \frac{\{ \psi_2, H_{R} \}}{2 \Phi}, \;\;
\lambda_2=- \frac{ \psi_2}{2 \Phi}\, .
\]
Thus
\begin{equation}
% \begin{split}
 \{j_i,   H\}=\{j_i,   H_{R}\}+ \lambda_\alpha
 \{j_i,\psi_\alpha \}= 
 \{j_i,   H_R\}- \sum_{\alpha, \beta=1}^2 \{j_i,\psi_\alpha \}D^{-1}_{\alpha \, \beta} 
 \{ \psi_\beta, H_{R} \}
=\{j_i,   H_{R}\}_D \, .
 \label{jiHRD}
%\end{split}
\end{equation}
Especially, it follows that
\[
\{\psi_\gamma,   H\}=  \{\psi_\gamma,   H_{R}\}- \sum_{\alpha,
\beta=1}^2 \{\psi_\gamma,
\psi_\alpha \}D^{-1}_{\alpha \, \beta} 
 \{ \psi_\beta, H_{R} \}=0,\qquad \gamma=1,2 \,.
 \]
 From equations \eqref{jiHRD} we obtain 
\begin{equation}
 \begin{split}
 \{j_i +j_{i+1},   H\}&=\{j_i+j_{i+1},   H_{R}\}
 - \{j_i+j_{i+1},\psi_1 \}D^{-1}_{1 \, 2} 
 \{ \psi_2, H_{R} \}-\{j_i+j_{i+1},\psi_2 \}D^{-1}_{2\, 1} 
 \{ \psi_1, H_{R} \} \\
 &=\{j_i+j_{i+1},   H\}
 - \{j_i+j_{i+1},\psi_2 \}D^{-1}_{2\, 1} 
 \{ \psi_1, H_{R} \} \, ,
 \label{jijipHD}
\end{split}
\end{equation}
where we used that 
\[\{j_i+j_{i+1},\psi_1 \}= -\{ j_i+j_{i+1}, \Phi\}=0\, ,\]   
ensuring that the cubic  term $\{ \psi_2, H_{R} \}$ never
appears in the dressing chain formulas.

After some algebra (and use
  of equation \eqref{j5zzero1} in the bottom equation below) we obtain:
\begin{equation}
\begin{split}
\{ j_1+j_{2} ,H \}& = -j_1^2+j_2^2 -\beta_1+\beta_2
+\frac{(j_1+j_2) \Psi}{\Phi}\, ,\\
\{ j_2+j_{3} ,H\}&=  -j_2^2+j_3^2 -\beta_2+\beta_3
-\frac{(j_2+j_3) \Psi}{\Phi}\, ,\\
\{ j_3+j_{4} ,H\} &= -j_3^2+j_4^2 -\beta_3+\beta_4 
+\frac{(j_3+j_4) \Psi}{\Phi}\, ,\\
\{ j_4+j_{1} ,H\}&= -j_1^2+j_4^2+2j_2^2-2 j_3^2 -\beta_1+2
\beta_2-2\beta_3\\&+\beta_4+
\frac{j_1+2j_2+2j_3+j_4}{\Phi} \Psi\\
&=
j_1^2-j_4^2 -\beta_4+
\beta_1 -\frac{(j_4+j_1) \Psi}{\Phi} \, .
\end{split}
 \label{j1234HRD}
\end{equation}
The above results \eqref{j1234HRD} can be summarized as
the result \eqref{main} for $N=4$.
Summing over $i=1,{\ldots} ,4$ gives \[ \{ \Phi ,H \} =0 \, ,\] which of
course does not contradict $\{ \Phi ,H_R \} =\Psi$ found earlier.

Also, one can form an alternating sum of expressions of 
both sides of \eqref{main} to obtain
\begin{equation}
\begin{split}
\sum_{i=1}^4 (-1)^i \{ j_i+j_{i+1} ,H \} &=
- \sum_{k=1}^4 \left( (-1)^k j_k^2 
+ \beta_k (-1)^k\right)+\sum_{k=1}^4 \left( (-1)^k j_{k+1}^2 
+ \beta_{k+1} (-1)^k\right)\\
&-\frac{f_1+f_2+f_3+f_4}{\Phi} \Psi = 2 \Psi-2 \Psi=0\, ,
\label{altersum}
\end{split}
\end{equation}
consistently with that the left hand side is identically zero for
$N$ even! 
Thus the
system of equations \eqref{main} does not need imposition of the condition
$\Psi=0$ for consistency.

Moreover, one can alternatively calculate 
\[
\{ j_i+j_{i+1}\, ,\, \frac13 \sum_{i=1}^4 j_i^3 + \sum_{i=1}^4 \beta_i
j_i\}_D\, ,\quad i=1,2,3,4\, ,
\]
with the Dirac bracket $\{ \cdot, \cdot \}_D$ as given in equations 
\eqref{taksolutions} instead of using bracket \eqref{Noddpb} and
Hamiltonian $H$ to reproduce the same result as in \eqref{j1234HRD}.

We now present a simple observation on how to explicitly 
transform any dressing chain of even cyclicity to the one with 
$\Psi = 0$ by shifting $j_i, i=1,{\ldots} , 2n$ by terms 
proportional to $\Psi/\Phi$.
Let us namely introduce
\begin{equation}
{\bar j}_{i}=  j_i +\frac{(-1)^{i}}{2} \frac{\Psi}{\Phi}
\label{jbardef}
\end{equation}
and notice that quantities $f_i = j_i+j_{i+1}=
{\bar j}_{i}+{\bar j}_{i+1}$ remain invariant under the shift in \eqref{jbardef}.
Accordingly, we can rewrite equation \eqref{main} 
(and its particular case in equation \eqref{j1234HRD}) as
\begin{equation} 
\{ {\bar j}_i+{\bar j}_{i+1} ,H \} = {\bar j}_{i+1}^2-{\bar j}_i^2+\beta_{i+1} -\beta_i
%+(-1)^{i+1} \frac{(j_i+j_{i+1}) }{\Phi}\Psi
, \quad i=1,2,{\ldots} ,N=2n \, , 
\label{barmain}
\end{equation}
removing explicit dependence on $\Psi$ and $\Phi$ in the commutation
relations. One can indeed check that 
${\bar  \Psi}=   {\bar j}_1^2+{\bar j}_3^2-{\bar j}_2^2-{\bar j}_4^2 
+\frac12 (-\alpha_1 +\alpha_2-\alpha_3+\alpha_4 )$ is identically zero 
for ${\bar j}_{i}$ defined in relation \eqref{jbardef} and so the system of equations \eqref{barmain} is
consistent.

Since $\{ j_i, \Psi\}_D=0$ 
for any $i$ the new quantities ${\bar j}_{i}$ will 
satisfy the same Dirac brackets \eqref{taksolutions} as the original
quantities $j_i$.

\subsection{Invariance under the  $A^{(1)}_{N-1}$ transformations}
\label{section:invariance}
In this subsection we show that equations
\eqref{main} exhibit invariance under the  $A^{(1)}_{N-1}$ transformations.
For illustration we here consider the $N=4$ example and 
$s_i,\, i=1,2,3,4$ transformations of $A^{(1)}_{3}$  \cite{adler}:
\begin{equation}
\begin{split}
 j_i & \stackrel{s_i}{\longrightarrow}
 \tilde{j}_i=  j_i - \frac{\kappa_i}{j_i
+j_{i+1}},\quad
j_{i+1}   \stackrel{s_i}{\longrightarrow}  \tilde{j}_{i+1}=  j_{i+1} + \frac{\kappa_i}{j_i
+j_{i+1}}, \\
j_k & \stackrel{s_i}{\longrightarrow} j_k, \; k \ne i, k\ne i+1\, ,
\end{split} \label{Tinewj}
\end{equation}
for $\kappa_i=\alpha_i=\beta_{i+1} - \beta_i$, when it is accompanied by transformations
of coefficients $ \alpha_i \to
-\alpha_i,  \alpha_{i \pm 1}\to \alpha_{i \pm 1} +\alpha_i$. This is
accomplished by the following $s_i$ transformation:
\begin{equation} 
\beta_i \stackrel{s_i}{\longrightarrow} \beta_{i+1}, \quad \beta_{i+1} 
\stackrel{s_i}{\longrightarrow}
\beta_i\, ,
\label{Tibetai}
\end{equation}

Explicit calculations show that the system of equations \eqref{main} is invariant
under the transformations \eqref{Tinewj} and  \eqref{Tibetai}.
The proof utilizes the fact that the objects $ \Psi, \Phi$ (and  obviously also
$f_i$ with the same index $i$ as in equation   \eqref{Tinewj}) 
are invariant under transformations
\eqref{Tinewj} and \eqref{Tibetai}. For $ \Psi$ this requires a small
calculation which for e.g. $s_1$ goes as follows:
\begin{equation}
\begin{split}
j_1^2-j_2^2+\beta_1-\beta_2 &\stackrel{s_1}{\to} \left( j_1 -
\frac{\kappa_1}{f_1}\right)^2-\left( j_2 +
\frac{\kappa_1}{f_1}\right)^2+\bar{\beta}_1-\bar{\beta}_2
= j_1^2-j_2^2- 2\kappa_1+\bar{\beta}_1-\bar{\beta}_2\\
&=j_1^2-j_2^2-2(\beta_{2} - \beta_1) +\beta_2-\beta_1= j_1^2-j_2^2+\beta_1-\beta_2 \, .
\label{Psis1invariance}
\end{split}
\end{equation}
Obviously  the $i-$th component of equations \eqref{main}
is invariant as $f_i$ remains unchanged. Explicitly, the
transformation of the right hand side of the $i-$th component of equation \eqref{main}
is as follows:
\[
\begin{split}
&-j_i^2+j_{i+1}^2 -\beta_i+\beta_{i+1}= f_i (j_{i+1}-j_i)
-\beta_i+\beta_{i+1}\\
&\stackrel{s_i}{\longrightarrow} (j_{i+1}+j_i)(j_{i+1}-j_i)+ 2 \kappa_i
\frac{f_i}{f_i} -\beta_{i+1}+\beta_i= (j_{i+1}+j_i)(j_{i+1}-j_i)
+\beta_{i+1}-\beta_i \, .
\end{split}
\]

However since $f_{i-1}$ (and $f_{i+1}$) is  are not invariant under
the transformation  \eqref{Tinewj} we need to separately consider  
the $(i+1)-$th (and the similar case of the $(i-1)-$th) component of 
equations \eqref{main} in order to explicitly
prove their invariance.
The left hand side becomes after $s_i$ transformation:
\[
\begin{split}
{\rm LHS}&= \{j_{i+1}+j_{i+2}, H\} - \kappa_i \frac{1}{(j_i+j_{i+1})^2}
\{j_{i+1}+j_{i+2}, H\}= \\
&=-j_{i+1}^2+j_{i+2}^2-\beta_{i+1}+\beta_{i+2}+(-1)^{i} \frac{f_{i+1}
}{\Phi}\Psi \\
&-\frac{\kappa_i}{f_i}(j_{i+1}-j_i)+\frac{\kappa_i(\beta_{i}-\beta_{i+1})}{f_i^2}
+(-1)^{i} \frac{\kappa_i}{f_i\Phi}\Psi \, ,
\end{split}
\]
while the right hand side becomes
\[
\begin{split}
{\rm RHS}&=
-(j_{i+1}+\frac{\kappa_i}{f_i})^2+j_{i+2}^2-s_i(\beta_{i+1})+s_i(\beta_{i+2})
+(-1)^i (f_{i+1}+\frac{\kappa_i}{f_i}) \frac{\Psi}{\Phi} = -j_{i+1}^2
-2 j_{i+1}\frac{\kappa_i}{f_i}\\
&-\frac{\kappa_i^2}{f_i^2}+j_{i+2}^2-\beta_i+\beta_{i+2}+(-1)^i
\frac{f_{i+1}+\kappa_i/f_i}{\Phi} \Psi \, .
\end{split}
\]
One can now easily show that the above two equations are equal to each other
using that $\kappa_i=\beta_{i+1}-\beta_i$ so that 
\[ -\frac{\kappa_i^2}{f_i^2}= \frac{\kappa_i(\beta_{i}-\beta_{i+1})}{f_i^2}
, \quad 
-2 j_{i+1}\frac{\kappa_i}{f_i}=-\frac{\kappa_i}{f_i}(j_{i+1}-j_i)
-\beta_{i+1}+\beta_i \, ,
\]
etc.

This establishes that the equations \eqref{main} are invariant under
B\"acklund transformations $s_i, i=1,{\ldots} , 4$.

Note that the automorphism $\pi$ such that $\pi(j_i)=j_{i+1},
\pi(\alpha_i)=\alpha_{i+1}$
transforms $\Phi, \Psi$ as follows
\[
\pi(\Phi) =\Phi, \qquad \pi(\Psi) =- \Psi
\]
and thus  the equations \eqref{main} are invariant under the
automorphism $\pi$ as
well, which completes the proof of  the  $A^{(1)}_{3}$ invariance.
Due to invariance of $\Phi, \Psi$ under Backlund symmetries it follows
easily that system of ${\bar j}_i, i=1,{\ldots} , 4$ defined in 
relation \eqref{jbardef} will transform under $s_i, \pi, i=1,{\ldots} , 4$ 
exactly as $j_i, i=1,{\ldots} , 4$.

The above result is consistent with the fact that 
the dressing equations \eqref{main}
can be cast in form of the Noumi-Yamada 
equations  \eqref{N4feqs} and these equations are known to be  invariant under 
the extended affine Weyl group. However for the dressing chain of even cyclicity  
there is no equivalent expression \eqref {fromf2j} 
that would give variables $j_i$ in terms of
variables $f_i$ of the Noumi-Yamada system. Thus the necessity 
to establish separately the invariance under $A^{(1)}_{N-1}$ transformations
for even dressing chains even more so because the invariance would not held for
the dressing chain \eqref{mape} with $N$ even but only for the equation 
\eqref{main} augmented by the $\Psi/\Phi$ terms.

\section{Deformation of the $N=4$ periodic dressing chain}
\label{section:deformation}
Having established the reduction from $N=5$ to $N=4$ periodic dressing
chains we will take advantage of the formalism to explore how to break
the extended affine Weyl symmetry $A^{(1)}_{3}$ symmetry by  explicit 
addition
of extra terms in the Hamiltonian.
We first perform a deformation of  $N=5$ symmetric Painlev\'e equations
\begin{equation}
f_{i\,, z}= f_i (f_{i+1}+f_{i-2}-f_{i+2}-f_{i-1}) +\alpha_i, \;
i=1,2,3,4,5\, .
\label{N5Noumi}
\end{equation}
Equations \eqref{N5Noumi}
are invariant under $s_i, i=1,2,3,4,5$ and $\pi$ transformations of the
extended affine $A^{(1)}_{4}$ Weyl group \cite{adler,noumi}.
Equations \eqref{N5Noumi} are also invariant under additional 
automorphisms 
\begin{equation}\begin{split}
\pi_i&: f_i \to - f_i , \;  f_{i-1} \to -f_{i+1}, \;  f_{i+1}, \to
-f_{i-1}, \;   f_{i-2} \to -f_{i+2}, \;
 f_{i+2} \to -f_{i-2},\\
 &:  \alpha_{i} \to-\alpha_i,\;\alpha_{i-1} \to -\alpha_{i+1}
 , \;  \alpha_{i+1} \to -\alpha_{i-1}, \;   \alpha_{i-2} \to
 -\alpha_{i+2}, \;
 \alpha_{i+2} \to -\alpha_{i-2},
 \label{N5pi0}
\end{split}\end{equation}
with $i=1,2,3,4,5$.

As the next step  we augment the Hamiltonian $H_5=\sum_{k=1}^5
(j_k^3/3 +\beta_k j_k )$  
by additional quadratic term: 
\begin{equation}
\begin{split}
H_{\rm deform}&=\frac12 \eta_1(f_2+f_3+f_4+f_5)^2 +
\frac12 \eta_2 (f_1+f_3+f_4+f_5)^2+
\frac12 \eta_3 (f_1+f_2+f_4+f_5)^2\\
&+\frac12 \eta_4 (f_1+f_2+f_3+f_5)^2
+\frac12 \eta_5 (f_1+f_2+f_3+f_4)^2\, ,
\label{Hdeformpi}
\end{split}
\end{equation}
with $\eta_i, i=1,{\ldots} , 5$ being deformation parameters.

Inserting $H=H_5+H_{\rm deform}$ into Hamilton equations leads to :
\begin{equation}\begin{split}
f_{i\,, z} &= f_i (f_{i+1}+f_{i-2}-f_{i+2}-f_{i-1})
+\alpha_i+\eta_{i+1} (\sum_{k\ne i+1} f_k) \\
&+ \eta_{i-3}  (\sum_{k\ne i-3} f_k) - \eta_{i+3}  (\sum_{k\ne i+3} f_k)
- \eta_{i-1}  (\sum_{k\ne i-1} f_k)\, ,
\label{N5deform}
\end{split}
\end{equation}
where e.g. $\sum_{k\ne 2} f_k =f_1+f_3+f_4+f_5$ and $i=1,{\ldots},5$.

Remarkably equations \eqref{N5deform} 
are still invariant under $\pi_i, i=1,{\ldots} ,5$
automorphisms \eqref{N5pi0} 
now extended to also act on the parameters $\eta_i$ as follows :
\begin{equation}%\begin{split}
\pi_i \,
  :\;  \eta_i \to -\eta_i, \;  \eta_{i-1} \to -\eta_{i+1}, \;  \eta_{i+1}, \to
-\eta_{i-1}, \;   \eta_{i-2} \to -\eta_{i+2}, \;
 \eta_{i+2} \to -\eta_{i-2}\, .
 \label{N5pi}
%\end{split}
\end{equation}
Equations \eqref{N5deform} are also invariant under the extended
automorphism $\pi: f_i \to f_{i+1}, 
\alpha_i \to \alpha_{i+1}, \eta_i \to \eta_{i+1}$ such that
$ \pi_4 \pi_3 \pi_2 \pi_1 = \pi^{-1}$.

However equations \eqref{N5deform} are not  symmetric under
B\"acklund transformations $s_i, i=1,{\ldots} ,5$ \eqref{Tinewj}. 
Setting some of the
parameters $\eta_i=0$ will, as we will see, restore invariance under
some of the  B\"acklund transformations $s_i$. 

We will use invariance under $\pi_i$ to guide us with regard to which
$\eta_i$ parameters to set to zero and proceed by choosing a model that is invariant under 
$\pi^2$ and $\pi_2$.
The $\pi_2$ invariant deformation is given by 
\begin{equation}
H_{\rm deform}^{(2)}=\frac12 \eta_1(f_2+f_3+f_4+f_5)^2 +
\frac12 \eta_2 (f_1+f_3+f_4+f_5)^2+
\frac12 \eta_3 (f_1+f_2+f_4+f_5)^2\, ,
\label{Hdeformpi2a}
\end{equation}
using the fact that $\pi_2: \eta_1 \to-\eta_3, \eta_3 \to -\eta_1, \eta_2\to-\eta_2$.
For $N=4$ reduction this gives:
\begin{equation}
H_{\rm deform}^{(2)}=\frac12 \eta_1(j_1+j_2)^2 +
\frac12 \eta_2(j_2+j_3)^2 +\frac12 \eta_3 (j_3+j_4)^2\, ,
\label{Hdeformpi2aN4}
\end{equation}
which maintains its invariance  under the reduction of $\pi_2$:
\begin{equation}
\begin{split}
{\hat \pi}_2&: j_2 \to-j_3,\; j_1 \to -j_4, \;  j_3 \to -j_2, \;   j_4 \to -j_1, \;
  \\
 &:  \beta_2 \to \beta_3,\;\beta_1 \to \beta_4
 , \;  \beta_3 \to \beta_2, \;   \beta_4 \to \beta_1, \;
\\
 &:  \eta_2 \to -\eta_2, \; \eta_1 \to - \eta_3, \; \eta_3 \to -\eta_1\, ,
\label{hatpi2}
\end{split}
\end{equation}
 Note that  $H=\sum_{k=1}^4 (j_k^3/3 +\beta_k j_k )
+H_{\rm deform}$  satisfies $\pi_i (H)=-H, i=1,{\ldots} ,5$. 
Further, setting $\eta_2=0$ we get 
$\pi_2 (H)=-H$ for $H=\sum_{k=1}^4 (j_k^3/3 +\beta_k j_k )
+H_{\rm deform}^{(2)} $, from which we obtain a system of 
equations (with $\{f,H\}$ replaced by $f_z$):
\begin{equation}
\begin{split}
( j_1 +j_{2})_z&= -j_1^2+j_2^2 -\beta_1+\beta_2+\frac{(j_1+j_2) \Psi}{\Phi}
 \\
( j_2 +j_{3})_z &=  -j_2^2+j_3^2 -\beta_2+\beta_3-\eta_1(j_1+j_2)
+\eta_3 (j_3+j_4) -\frac{(j_2+j_3) \Psi}{\Phi}
 \\
( j_3 +j_{4})_z&=  -j_3^2+j_4^2 -\beta_3+\beta_4 +\frac{(j_3+j_4) \Psi}{\Phi}
 \\
( j_4 +j_{1})_z&= -j_1^2+j_4^2+2j_2^2-2 j_3^2 -\beta_1+2
\beta_2-2\beta_3+\beta_4\\
&+\eta_1(j_1+j_2)-\eta_3 (j_3+j_4)+
\frac{j_1+2j_2+2j_3+j_4}{\Phi} \Psi\, ,
\label{jN4deform2b}
\end{split}
\end{equation}
that will not only maintain the $\pi_2$ symmetry from \eqref{hatpi2} 
but also be invariant under $s_1,s_3$ transformations. It will also be invariant 
under  $\pi^2$ extended by $\pi^2: \eta_1 \to
\eta_3$ and under ${\hat \pi}_2$ from \eqref{hatpi2}.
%(after introducing the
%Dirac bracket with explicit presence of the $\Psi$ constraint).

The symmetry operations of equations \eqref{jN4deform2b} satisfy the relations
\[
{\hat \pi}_2 s_1 =s_3 {\hat \pi}_2, \; \pi^2 s_1 =s_3 \pi^2
, \; \pi^2 {\hat \pi}_2={\hat \pi}_2 \pi^2
\]
where $s_1,s_3$ were defined in equations \eqref{Tinewj}.

The constraint $\Psi$ in equation \eqref{jN4deform2b} 
does not depend on $\eta$'s and will satisfy:
$ {\hat \pi}_2 (\Psi)= - \Psi$ and $\pi^2 (\Psi)= \Psi$.
As in equation \eqref{N4feqs} we can rewrite equations \eqref{jN4deform2b}
in a compact way
\[
\Phi\, f_{i\, ,z}=f_i f_{i+2} \left(f_{i+1}-f_{i-1}\right) +
(-1)^i \left(\alpha_1+\alpha_3 \right) f_i + \alpha_i (f_i+f_{i+2})+
(\delta_{i,2}-\delta_{i,4})(-\eta_1 f_1+\eta_3 f_3)
\]
for $i=1,2,3,4$.

We recognize in the above equations a model proposed earlier
\cite{p3-p5} within the framework  of Painlev\'e equations and
shown to
pass the Painlev\'e test due to the presence of the 
remaining B\"acklund symmetries. Generally,  the deformed 
models studied in the literature, \cite{victoretal,p3-p5}, 
pass the Painl\'eve test if the deformation
maintains invariance under at least one of the $s_i, i=1, ,2, ...,
N-1$ of the original B\"acklund transformations but fail to pass the 
test if all of $s_i, i=1, ,2, ..., N-1$ are broken.
The model described by equations \eqref{jN4deform2b} is invariant
under the two B\"acklund symmetries $s_1, s_3$. Since it is fully equivalent to the model
considered in \cite{victoretal} in the setting of Painlev\'e equations it falls 
within a class of models that pass the Painlev\'e test.

\section{Connection to $2n$ boson models}
\label{section:2011paper}

In \cite{2010paper,2011paper} we proposed self-similarity reductions
of the constrained KP hierarchy with symmetry structure defined
by B\"acklund transformations induced by a discrete structure of 
Volterra type lattice. The models and their self-similarity reductions were 
referred to as ``$2n$ boson models''.
 
The $2n$ boson models models were conveniently expressed in \cite{2011paper}
in terms of $n$ canonical pairs 
$e_i, Y_i, i=1,2, {\ldots} , n$ satisfying the  bracket 
$\{ e_i , Y_j \} = \delta_{ij}, \; i=1, {\ldots}, n,$
which enter the Hamiltonian (here for simplicity  given for $n=2$):
\[
{\cal H}_{A^{(1)}_{4}} =- \sum_{j=1}^2 e_j  \left(Y_j -2x \right) \left(Y_j -e_j\right)
+2  e_1\left(Y_1 -2x \right) \left(Y_2-e_2\right) %\nonumber\\
+\sum_{j=1}^2 {\bar k}_j Y_j 
- \sum_{j=1}^2    {\kappa}_j  e_j \,
\]
The reference  \cite{2011paper} presented an explicit symplectic map
from canonical variables 
\[p_i=f_{2i}, \quad q_i=\sum_{k=1}^i f_i,\quad
i=1,2,{\ldots} ,n\] 
from $A^{(1)}_{2n}$ Noumi-Yamada system \cite{noumi} in terms of $f_i$ to the $2n$
boson model represented by $e_i, Y_i, i=1,2, {\ldots} , n$ variables. 
This construction has established  equivalence of $2n$ boson model 
to symmetric Painlev\'e systems with $A^{(1)}_{2n}$
Weyl symmetry 
and therefore also equivalence to $N=2n+1$ periodic dressing chain systems.

To derive the reduced system invariant under
$A^{(1)}_{2n-1}$
Weyl symmetry the paper \cite{2011paper}  adopted the Dirac technique
with the primary constraint:
\begin{equation}
\psi_1=Y_2=0\, .
\label{psi1Y2}
\end{equation}
In what follows we compare  the  approach of  \cite{2011paper} to the method 
proposed in this paper. 

The Dirac procedure of  \cite{2011paper} went as follows.
Once we set the primary constraint \eqref{psi1Y2},
the secondary constraint $\psi_2$ then follows from
\begin{equation}
\psi_2=-\frac{\partial {\cal H}_{A^{(1)}_{4}}}{ \partial
e_2}= 4 x e_2 +2 e_1 (Y_1-2 x)+ {\kappa}_2 \, .
\label{psi2e2}
\end{equation}
The fundamental bracket is
\begin{equation}
\{ \psi_1, \psi_2 \}= -4 x \ne 0\, ,
\label{psi1psi2a}
\end{equation}
and can be used to calculate the Dirac brackets:
\begin{equation}
\{ e_2, e_1 \}_D= \frac{e_1}{2x}, \quad \{e_2 , Y_1\}_D=
-\frac{Y_1-2x}{2x}\, .
\label{e2e1}\end{equation}
The other bracket $\{e_1,Y_1\}=\{e_1,Y_1\}_D=1$ is unchanged. It follows 
from the Dirac brackets \eqref{e2e1} that
\begin{equation}
\{ e_1, \psi_\alpha \}_D= 0, \quad \{Y_1, \psi_\alpha\}_D=
0\qquad \alpha=1,2\, .
\label{e1Y1psi}\end{equation}
We can therefore  directly implement reduction by substituting
\[Y_2=\psi_1,\quad e_2= \left(\psi_2-2 e_1 (Y_1-2x)-\kappa_2\right)/4x\]
into ${\cal H}_{A^{(1)}_{4}} $  to obtain the reduced Hamiltonian :
\begin{equation} 
%\begin{split}
2 x {\bar {\cal H}}_2  =  e_1  \left(Y_1 -2x \right)
Y_1 \left( e_1-2 x \right)
+ {\kappa}_2 e_1 Y_1
+ {\bar k}_1 2 x Y_1 \\
-   \left( {\kappa}_1+{\kappa}_2 \right) 2 x e_1\, ,
%\end{split}
\label{hr2eY}
\end{equation}
where for simplicity we set  $\psi_1=0, \psi_2=0$, since they
both commute with $e_1,Y_1$.
The corresponding Hamilton equations are :
\begin{equation}
\begin{split}
\{ e_1, {\bar {\cal H}}_2 \}
&= 2 x e_1-e^{2}_{1}
-2 e_{1} Y_1 +{\bar k}_1
+\frac{1}{2x} \left( 2e^{2}_{1}Y_1  +e_1 \kappa_2\right)\\
\{ Y_1, {\bar {\cal H}}_2 \} &= -2x Y_1 +2 e_1Y_1+Y_1^2 + \kappa_1+\kappa_2
-\frac{1}{2x} \left( 2Y_1^2 e_1+ Y_1 \kappa_2\right) \, .
\label{e1Y}
\end{split}
\end{equation}
In what follows we will show their complete equivalence to equations \eqref{main} for
$N=4$.

Mapping of $N=5$ brackets for $f_i$ into the $e_1,Y_1,e_2,Y_2$ 
goes through the canonical $p,q$ system :
\[
p_1= f_2, \; p_2=f_4, \; q_1=f_1, \; q_2= f_1+f_3
\]
such that $\{ q_i , p_j\}=\delta_{ij}$. The symplectic map 
from  $p_i,q_j$  to $e_i,Y_j $ is as follows:
 \begin{equation}
 \begin{split}
 Y_2&=q_1+p_2+2x = f_1+f_4+2x , \quad
 e_2=q_2+p_2+2 x =f_1+f_3+f_4+2 x \\
 Y_1&=-q_2-p_2-p_1= -f_1- f_3-f_2 -f_4, \quad e_1=-q_1=-f_1\, .
 \label{schemeqp}
 \end{split}
 \end{equation}
We now impose the constraint $\psi_1=Y_2$ followed by 
the secondary constraint $\psi_2=4 x e_2+2
e_1(Y_1-2x)+\kappa_2$ and show that this is equivalent 
to $N=4$ model obtained by reducing the Shabat-Veselov's $N=5$ dressing model.

From the top equation of \eqref{schemeqp} and the fact that
\[
-2 x = f_1+f_2+f_3+f_4+f_5= 2 (j_1+j_2+j_3+j_4+j_5)
\]
for the $N=5$ system it follows that 
\[
%\begin{split}
Y_2 =  f_1+f_4 -(f_1+f_2+f_3+f_4+f_5)=-j_1-j_2-2j_3-j_4-j_5=-j_3+x
%\end{split}
\]
Thus $Y_2=0$ is equivalent to $j_3=x$ or
\begin{equation}
Y_2=0 \, \longrightarrow \, j_1+j_2+j_4+j_5=-2x\, .
\label{Y20Phi}
\end{equation}
Based on this relation we now can define the $N=4$ system as
\begin{equation}
 \begin{split}
{\tilde f}_1&= j_1+j_2, \quad {\tilde f}_3=j_4+j_5 ={\tilde
j}_3+{\tilde j}_4, \\
{\tilde f}_2&=j_2+j_4= j_2+{\tilde
j}_3, \quad {\tilde f}_4=j_5+j_1={\tilde j}_4+j_1,
 \label{tildef}
 \end{split}
 \end{equation}
where we have introduced 
\begin{equation}
{\tilde j}_3= j_4, \quad {\tilde j}_4 = j_5\, ,
\label{tildej}
\end{equation}
such that
\begin{equation}
{\tilde f}_1+ {\tilde f}_3=-2x, \quad
{\tilde f}_2+ {\tilde f}_4=-2x\, ,
\label{tildef2x}
\end{equation}
equivalent to
\begin{equation}
j_1+j_2+ {\tilde
j}_3+{\tilde j}_4=-2x\, .
\label{tildej2x}
\end{equation}
To calculate the secondary constraint $\psi_2$ we need to express 
$e_1,e_2,Y_1$ in terms of $N=4$ quantities that is as follows:
\begin{equation}
 \begin{split}
e_1&= - f_1=-j_1-j_2\\
e_2&=q_2+p_2+2 x =f_1+f_3+f_4+2 x = j_1+j_2+j_3+2 j_4+j_5+2x\\
&=-j_1/2-j_2/2+{\tilde j}_3/2-{\tilde j}_4/2\\
Y_1&= -f_1- f_3-f_2 -f_4=-j_1-2j_2-2j_3-2j_4-j_5
=- j_2- {\tilde j}_3=-{\tilde f}_2\, .
\label{e1e2Y1}
\end{split}
 \end{equation}
In the above equation we used identities that hold for $Y_2=0$.

We can now calculate the secondary constraint $\psi_2$:
\begin{equation}
 %\begin{split}
\psi_2=4 x e_2+2
e_1(Y_1-2x)+\kappa_2 = 2 \lbrack 2 x e_2+e_1(Y_1-2x)\rbrack +\kappa_2 
= -j_1^2+j_2^2-{\tilde j}_3^2+{\tilde j}_4^2  +\kappa_2 \, .
\label{psi2proof}
% \end{split}
 \end{equation}
Recovering our constraint $\Psi$ from section \ref{section:reduction}
up to an overall sign. Thus the Dirac reduction from
 \cite{2011paper} agrees with the Dirac reduction we have designed for
 the periodic dressing chain equations.
 
On basis of identification \eqref{psi2proof} we can 
rewrite the first of Hamilton equations \eqref{e1Y} as :
\begin{equation}
\begin{split}
\{ e_1, {\bar {\cal H}}_2 \} &= 2 x e_1-e^{2}_{1}
-2 e_{1} Y_1 +{\bar k}_1
+\frac{e_1}{2x} \left( 2e_{1}Y_1  + \kappa_2+4x e_2 -4 xe_2 -4 x e_1 +
4x e_1 \right)\\
&=j_1^2-j_2^2 +{\bar k}_1 +\frac{e_1}{2x} \psi_2\, ,
\label{e1x}
\end{split}
\end{equation}
after we inserted values of $e_1,Y_1,e_2$ in terms of $j_1,j_2,{\tilde
j}_3, {\tilde j}_4$ from equations \eqref{e1e2Y1}. 

Recalling from  equation \eqref{e1e2Y1} that $e_1= -j_1-j_2$ and
inserting it into the equation \eqref{e1x} we find
\begin{equation}
%\begin{split}
\{ j_1+j_2, {\bar {\cal H}}_2 \} 
= j_2^2-j_1^2 -{\bar k}_1+\frac{j_1+j_2}{-2x} \Psi
=j_2^2-j_1^2 +\alpha_1+\frac{j_1+j_2}{-2x} \Psi\,,
\label{works}
%\end{split}
\end{equation}
where $\alpha_1=-{\bar k}_1$,
$-2 x = \Phi$ and $\Psi=-\psi_2$ to agree with our convention.
We recognize in \eqref{works} the first dressing equation of
\eqref{j1234HRD}.

Let us rewrite the second of Hamilton equations \eqref{e1Y} as :
\begin{equation}
\begin{split}
\{ Y_1, {\bar {\cal H}}_2 \} &= -2x Y_1 +2 e_1Y_1+Y_1^2 + \kappa_1+\kappa_2
-\frac{1}{2x} \left( 2Y_1^2 e_1+ Y_1 \kappa_2\right) \\
&=j_2^2-{\tilde j}_3^2+ \kappa_1+\kappa_2
-\frac{Y_1}{2x} \, \psi_2\, .
\label{Y1x}
\end{split}
\end{equation}
Recalling that $Y_1=-(j_2+{\tilde j}_3)$ we can rewrite the above as :
\begin{equation}
%\begin{split}
\{ j_2+{\tilde j}_3, {\bar {\cal H}}_2 \}= {\tilde j}_3^2-j_2^2 -\kappa_1-\kappa_2
-\frac{j_2+{\tilde j}_3}{-2x} \, \Psi
={\tilde j}_3^2-j_2^2 +\alpha_2
-\frac{j_2+{\tilde j}_3}{-2x} \, \Psi\, ,
\label{j2tj3}
%\end{split}
\end{equation}
where we set $\alpha_2= -\kappa_1-\kappa_2$, $-2 x = \Phi$ and $\Psi=-\psi_2$ to agree with our earlier
convention.
We recognize in \eqref{j2tj3} the second dressing equation of
\eqref{j1234HRD}.

Using relation \eqref{tildej2x} we can similarly derive equations
for $\{ j_1+{\tilde j}_4,  {\bar {\cal H}}_2 \}$ and 
$\{{\tilde j}_3+{\tilde j}_4,  {\bar {\cal H}}_2 \}$
obtaining all the dressing equations \eqref{j1234HRD} 
from the $e$-$Y$ system.

Repeating almost verbatim what we have done in reference \cite{2011paper} 
we can cast equations \eqref{e1Y} into into 
symmetric $A^{(1)}_3$ Painlev\'e V equations \eqref{N4feqs}
using that  from relation \eqref{e1e2Y1}
it follows that  $Y_1=-f_2$ and 
$e_1=- f_1$. Subsequently $f_4=-2x+Y_1$ and
$f_3=-2x+e_1$ (we we ignore tildes for simplicity).
As defined above the constants are :
\[\alpha_1=-{\bar k}_1,\;\;\; \alpha_2=-\kappa_1-\kappa_2,\;\;\; 
\alpha_3=\kappa_2+{\bar k}_1, \; \alpha_4 =\kappa_1 \, .
\]

\section{Conclusion and outlook}
We proposed a consistent and systematic approach based on Dirac
reduction framework
to formulating  dressing chain equations  for even 
$2n$ periodicity. Such approach  is naturally obtained  from 
the corresponding dressing chain equations of odd $2n+1$
periodicity by reduction. Both chains of even or odd 
periodicity are equivalent to 
$A^{(1)}_{2n-1}$, $A^{(1)}_{2n}$ Painlev\'e systems, respectively. 
The formalism is in agreement with the previous result obtained in a
setup of $2n$ boson models  \cite{2011paper}
and facilitates further studies of deformations 
of the original B\"acklund symmetry.

Among subjects deserving further investigation is establishing the bi-Hamiltonian
nature of the  dressing chains with even 
periodicity. The bi-Hamiltonian character of the  dressing chains with odd
periodicity is well known \cite{veselov} and in further development 
separation of variables has been developed in such case \cite{blaszak}. 
Uncovering the bihamiltonian
structure for  dressing chains of even cyclicity and related development of 
separation of variables is a proposal to be studied elsewhere.

\subsection*{Acknowledgements}

JFG and AHZ thank CNPq and FAPESP for partial  support. 
We thank the referee for useful comments towards improving the
manuscript.

\label{lastpage}

\end{document}